\documentclass[aps,showpacs,prd,twocolumn,10pt,floatfix,nofootinbib]{revtex4}

\usepackage[latin1]{inputenc}
\usepackage{bbm}
\usepackage{bm}
\usepackage{graphicx}
\usepackage{epsfig}
\usepackage{subfigure}
\usepackage{latexsym}
\usepackage{amsmath}
\usepackage{amsfonts}
\usepackage{amssymb}
\usepackage{color}
\usepackage{hyperref}
\usepackage{longtable}

\newcommand{\Tr}{\textrm{Tr}}

\newcommand{\ev}[1]{\left\langle #1 \right\rangle}
\newcommand{\dd}{\textrm{d}}

\newcommand{\eq}[1]{Eq.~(\ref{#1})}

\newcommand{\be}{\begin{eqnarray}} 
\newcommand{\ee}{\end{eqnarray}}

\newcommand{\bmp}{\noindent\begin{minipage}{16cm}}
\newcommand{\emp}{\end{minipage}\vskip 7mm} 
\def\lsim{\mathrel{\raise.3ex\hbox{$<$\kern-.75em\lower1ex\hbox{$\sim$}}}}
\def\gsim{\mathrel{\raise.3ex\hbox{$>$\kern-.75em\lower1ex\hbox{$\sim$}}}}

\begin{document}


\title{Evolution of the coupling constant in SU(2) lattice gauge 
  theory with two adjoint fermions}

\author{Ari J. {\sc Hietanen}}
\affiliation{Department of Physics, Florida International University}
\author{Kari {\sc Rummukainen}}
\affiliation{Department of Physics and 
  Helsinki Institute of Physics,
  University of Helsinki, Finland}
\author{Kimmo {\sc Tuominen}}
\affiliation{Department of Physics, University of Jyv\"askyl\"a and 
  Helsinki Institute of Physics, University of Helsinki, Finland}

\begin{abstract}
  We measure the evolution of the coupling constant using the
  Schr\"odinger functional method in the lattice formulation of SU(2)
  gauge theory with two massless Dirac fermions in the adjoint representation.
  We observe strong evidence for an infrared fixed point, where
  the theory becomes conformal.  We measure the 
  $\beta$-function and the coupling constant as a function of 
  the energy scale.
\end{abstract}


\maketitle


\section{Introduction}

In preparation for the phenomenology at LHC, there
has recently been increased interest in field theories which appear
(quasi) conformal when probed towards infrared, i.e. over large distance
scales.  There are basically two novel applications which utilize such
theories, namely walking technicolor \cite{Hill:2002ap} and unparticles
\cite{Georgi:2007ek,Sannino:2008nv}. For model building
along these directions it is very desirable to single out candidate
theories which allow for the existence of a nontrivial infrared fixed point and
then use these as a basis for more detailed analyses. 
To determine the location of the conformal window as a function of the number of colors, flavors and fermion representations is a nonperturbative problem. Traditional way to obtain semi-quantitative estimates is  the so-called rainbow approximation to the Schwinger-Dyson equation for the nonperturbative fermion propagator \cite{Pagels:1974se,Fukuda:1976zb,Appelquist:1988yc}. More recent developments include analytic $\beta$-function ans\"atze \cite{Ryttov:2007cx,Dietrich:2009ns,Antipin:2009wr} and analyses on the role of topological excitations \cite{Poppitz:2009uq}. Phenomenological constraints on walking technicolor require that conformal behavior is obtained with only modest amount of new matter fields, and utilizing higher representation fermions, several plausible candidates have been proposed in \cite{Sannino:2004qp,Ryttov:2008xe}. Although subject to uncertainties, these methods are invaluable in providing hints towards theories which are of phenomenological interest and motivate their analyses using lattice methods.  

Guided by the phase diagrams of \cite{Sannino:2004qp}, initial lattice studies of two and three color theories with two fermion flavours in the two-index symmetric representation have recently appeared \cite{Catterall:2007yx,Shamir:2008pb,DelDebbio:2009fd,Fodor:2009ar,Bursa:2009tj,Sinclair:2009ec}. For related studies with fundamental fermions in SU(3) gauge theory, see \cite{Appelquist:2007hu,Fodor:2009wk,Deuzeman:2009mh,Jin:2009mc,Hasenfratz:2009ea}.  

In this work we study the evolution of the coupling constant of the
candidate model with ``minimal'' field content, consisting of the
standard SU(2) gauge field and two massless Dirac fermions in the
adjoint representation.  
The euclidean Lagrange density is
\begin{equation}
  \mathcal{L} = \frac12 \Tr F_{\mu\nu}F_{\mu\nu} 
  + \mbox{$\sum_f$} \bar\psi_f i \gamma_\mu D_\mu \psi_f
\end{equation}
with $D_\mu = \partial_\mu + i g A^a T^a$, where $(T^a)^{bc} =
-i\epsilon^{abc}$ are the 
generators of the adjoint representation.
To obtain the evolution of the couplig we use the lattice 
Schr\"odinger functional scheme, which 
has been very successfully applied to e.g. two-flavour QCD by the
Alpha collaboration \cite{DellaMorte:2004bc}. 

Perturbatively, the two-loop $\beta$-function of this theory with
$N_f$ adjoint fermions is
\begin{equation}
  \beta(g)=\mu \frac{\dd g}{\dd\mu}=-\beta_0\frac{g^3}{16\pi^2}
  -\beta_1\frac{g^5}{(16\pi^2)^2},
  \label{pert_beta}
\end{equation}
where $\beta_0=(22-8N_f)/3$ and $\beta_1=4(34-32N_f)/3$. For $N_f=1,2$
the coefficient $\beta_0$ is positive while $\beta_1$ is positive for
$N_f=1$ but negative for $N_f=2$. Hence, for $N_f=2$ there is a zero
of the two-loop $\beta$-function at 
$g_{\ast}^2 = -(\beta_0/\beta_1)16\pi^2 \approx 7.9$, 
i.e. $\alpha_\ast=g_{\ast}^2/(4\pi) \approx 0.6$,  
corresponding to an infrared fixed point (IRFP).
Large value of $g_\ast^2$ casts
doubt on the validity of the perturbative result and nonperturbative
measurement of the running coupling on the lattice is required.
Indeed, as we report below, the lattice simulations reveal large deviations
from the two-loop perturbative behaviour.

We note that the $\beta$-function has been calculated up to four loops in the Minimal Subtraction (MS) scheme \cite{vanRitbergen:1997va}.  Beyond two loops the $\beta$-function is scheme dependent and the result cannot be directly compared with the lattice Schr\"odinger functional scheme used here.

For the gauge group SU(2) considered here the two-index symmetric representation coincides with the adjoint representation. We have recently performed large-volume high-statistics lattice simulations of this theory, concentrating on the chiral properties of the mass spectrum \cite{Hietanen:2008mr,Hietanen:2008vc}. The main result of this study was that at small enough lattice gauge coupling, in contrast to QCD, 
we did not observe spontaneous chiral symmetry breaking as the quark mass was decreased.  Because the chiral condensate generates a mass scale, its existence would exclude the IRFP.  While the non-observation of the spontaneous chiral symmetry breaking is very suggestive, it by no means proves that the IRFP exists at non-zero coupling, and one needs to measure the evolution of the coupling directly \footnote{Alternatively one can study the properties of the physical degrees of freedom in the limit of zero quark mass as outlined in \cite{Sannino:2008pz}.}.

\section{Lattice formulation}
The lattice theory is defined with the action
\begin{equation}
S = S_\mathrm{G} + S_\mathrm{F},
\end{equation}
where $S_\mathrm{G}$ is the standard Wilson plaquette action for SU(2)
gauge fields:
\begin{equation}
  S_\mathrm{G} = \beta_L \sum_{x,\mu<\nu} \left( 1 - 
  \frac12 \Tr [U_{x,\mu}U_{x+\hat\mu,\nu} 
  U^\dagger_{x+\hat\nu,\mu} U^\dagger_{x,\nu}] \right).
\end{equation}
$S_\mathrm{F}$ is the Wilson fermion action 
for two Dirac fermions in the adjoint representation of SU(2):
\begin{equation}
  S_\mathrm{F} =
  \sum_{f=\textrm{u},\textrm{d}} 
  \sum_{x,y} \bar\psi_{f,x} M_{xy} \psi_{f,y} ,
  \label{sf}
\end{equation}
where 
\begin{equation}
  M_{xy} = \delta_{xy} - 
  \kappa \sum_\mu \left[ (1+\gamma_\mu) V_{x,\mu} +
    (1-\gamma_\mu)V^T_{x-\mu,\mu}\right].
\end{equation}
It differs from the standard Wilson fermion action only by the
replacement of the fundamental representation link matrices
$U_\mu(x)$ with the adjoint representation ones:
\begin{equation}
  V_\mu^{ab}(x)=2 \Tr(\lambda^aU_\mu(x)\lambda^bU_\mu^\dagger(x)),
\end{equation}
where $\lambda^a=\frac12 \sigma^a$, $a=1,2,3$, are the 
generators of the fundamental representation. 
The lattice action is conventionally 
parametrised with two dimensionless parameters,
\begin{equation}
  \beta_L = \frac{4}{g_0^2},
  ~~~~~~~
  \kappa = \frac{1}{8+2 am_{q,0}},
\end{equation} 
where $a$ is the lattice spacing,
and $g_0^2$ and $m_{q,0}$ the bare gauge coupling and quark
mass.  Because the Wilson action explicitly breaks the
chiral symmetry of the continuum quark action, quark masses
are additively renormalised and the bare quark mass must be
tuned to achieve the massless limit.
In order to keep the methodology simple, we do not implement
$O(a)$ improvement on the lattice.

In earlier studies \cite{Hietanen:2008mr,Catterall:2007yx} it has been
found that in this lattice theory there exists a critical coupling
$\beta_{L,c} \approx 2$ so that when $\beta_L < \beta_{L,c}$ the
theory shows signs of spontaneous chiral symmetry breaking, 
but reaching the limit $m_Q\rightarrow 0$ is prevented 
by the appearance of a first order transition where $m_Q$ jumps to negative values.  
However, when $\beta_L > \beta_{L,c}$ the $m_Q\rightarrow 0$ limit can be reached without 
observing the spontaneous chiral symmetry breaking. This is the region where it is possible to find conformal behaviour in the limit of massless fermions. 

Because the Wilson fermion action breaks the continuum chiral symmetry,
the quark mass is additively renormalised and we determine the quark mass 
through the PCAC relation
\begin{equation}
  2 m_Q = \left[ {\partial_t\langle V_A(x) \rangle} / 
    {\langle V_P(x) \rangle}\right]_{\vert_{x_0=L/2}}
\label{mq}
\end{equation}
where $V_A$ and $V_P$ are (non-improved) axial and pseudoscalar
current correlators \cite{Luscher:1996ug}:
\begin{align}
  V_A(x) &= a^6 \sum_{\bm{y},\bm{z}} 
  \ev{
    \bar{\textrm{d}}(x)\gamma_0\gamma_5 \textrm{u}(x)\,
    \bar{\textrm{u}}(\bm{y},0)\gamma_5 \textrm{d}(\bm{z},0)
  } \\
  V_P(x) &= a^6 \sum_{\bm{y},\bm{z}} 
  \ev{ 
    \bar{\textrm{d}}(x)\gamma_5 \textrm{u}(x)\,
    \bar{\textrm{u}}(\bm{y},0)\gamma_5 \textrm{d}(\bm{z},0)
  }
\end{align}
Here d and u refer to two quark flavours, and 
at time slices $x_0=0$ and $x_0=L$ the gauge
fields are fixed to specific Schr\"odinger function
boundary values described in detail below.  
For each lattice coupling $\beta_L$ we tune
$\kappa$ so that $m_Q$ vanishes. This determines the critical line
$\kappa_c(\beta_L)$; from now on we assume that we always tune to
the limit $m_Q=0$.

The coupling is measured using the Schr\"odinger functional (SF) method 
\cite{Luscher:1991wu,Luscher:1992an,Sint:1995rb}, 
i.e. introducing a background field using special boundary conditions
and studying the response of the system to changes of the background
field. The scale at which the coupling is measured is determined by
the finite size of the system.  Here we consider lattices of volume 
$V=L^4=(Na)^4$, and following \cite{Luscher:1992zx}, the
spatial gauge links on the $x_0=0$ and $x_0=L$ boundaries are fixed to 
constant diagonal SU(2) matrices:
\begin{eqnarray}
  U_\mu(x_0 = 0) &=& \exp(-i\eta\sigma_3 a/L)\\
  U_\mu(x_0 = L) &=& \exp(-i(\pi-\eta)\sigma_3 a/L),
\end{eqnarray}
where $\sigma_3$ is the third Pauli matrix. 
The spatial gauge field boundary conditions are periodic.
During the simulation the fermion fields are set to vanish at 
boundaries $x_0=0$ and $x_0=L$,
and are ``twisted'' periodic to spatial directions:
$\psi(x+L\hat e_i) = \exp(i\pi/5)\psi(x)$ \cite{Sint:1995rb}. 

At the classical level the gauge field boundary conditions 
generate a constant color diagonal chromoelectric field, 
and the derivative of the action
with respect to $\eta$ is easily calculable:
\begin{equation}
  \frac{\partial S^{\textrm{cl.}}}{\partial \eta} =
  \frac{k}{g_{0}^2}\,,
\end{equation}
where, on a lattice of size $L^4 = (Na)^4$ \cite{Luscher:1992zx},
\begin{equation}
  k = -24 (L/a)^2 \sin\frac{\pi-2\eta}{(L/a)^2}.
\end{equation}
At quantum level the boundary conditions generate
a  background field.  Using
an effective action $\Gamma$,
\begin{equation}
  e^{-\Gamma}=\int{\mathcal{D}}[U,\psi,\bar{\psi}]e^{-S},
\end{equation}
the coupling constant is defined through
\begin{equation}
  \frac{\partial \Gamma}{\partial \eta} = 
  \left\langle\frac{\partial S_{\textrm{G}}}{\partial\eta}\right\rangle =
  \frac{k}{g^2(L)}\,.
\end{equation}
We fix $\eta = \pi/4$ after taking the derivative.
For our action the observable 
$\langle \partial S_{\textrm{G}} / \partial\eta \rangle$
is proportional to the expectation value of the boundary plaquette
and easily measurable.
%
The result is the coupling constant $g^2(L)$ as a function of the physical size of the lattice.  Because a priori the lattice spacing $a$ is unknown, we denote the lattice measurements with $g^2(L/a,\beta_L)$. Formal continuum limit is obtained by keeping $g^2$ constant while taking $L/a\rightarrow\infty$.

\section{Results and analysis}

As stated in the previous section we will consider lattice couplings
$\beta_L < \beta_{L,c}\approx 2$ and tune to the $m_Q=0$ limit
corresponding to a critical value $\kappa_c(\beta_L)$. The values of
$\kappa_c$, as well as the corresponding $m_Q$, at each value of
$\beta_L$ used in this study are shown in Table \ref{tab:par}. The
$\beta_L$-values which we consider are all larger than the critical
value $\beta_{L,c}\approx 2$; indeed, already at $\beta_L = 1.9$ it
was impossible to reach $m_Q = 0$ limit.

\begin{table}
\centerline{
\begin{tabular}{|@{~~}l@{~~~}|@{~~~}l@{~~~}|@{~~~}l@{~~~}|}
\hline
$\beta_L$ & $\kappa_c$ & $m_Q a$ \\
\hline
 2.05    & 0.18625  & -0.00377(29) \\
 2.2     & 0.1805   &  0.00016(16) \\
 2.5     & 0.17172  & -0.00079(10) \\
 3       & 0.161636 &  0.00213(11) \\
 3.5     & 0.155132~&  0.00028(11) \\
 4.5     & 0.14712~ &  0.00000(05) \\
 8       & 0.136415~& -0.00038(04) \\
\hline
\end{tabular}}
\caption{Values of $\beta_L$ and the estimates of $\kappa_c$ used. 
$\kappa_c$ is determined on $16^4$ lattices with small statistics runs.
The table also shows the small residual value of $m_Q a$ measured from the
much larger statistics production runs on $16^4$ lattices.}
\label{tab:par}
\end{table}

\begingroup
\squeezetable
\begin{table}
\centerline{
\begin{tabular}{|@{~~~}l@{~~~}|@{~~~}r@{~~~}|@{~~~}r@{~~~}|@{~~~}l@{~~~}|}
\hline
$\beta_L$ & $N$ & $N_{\rm traj.}$ & $1/g^2$ \\
\hline  
2.05 & 4 & 20000 & 0.2504(17) \\
    & 6 & 51200 & 0.2233(16) \\
    & 8 & 29000 & 0.2179(36) \\
    & 12 & 27000 & 0.2187(33) \\
    & 16 & 30900 & 0.2391(53) \\
    & 20 & 17300 & 0.257(22) \\
\hline
2.2 & 4 & 16000 & 0.3069(17) \\
    & 6 & 51200 & 0.2857(15) \\ 
    & 8 & 19900 & 0.2719(36) \\
    & 12 & 55000 & 0.2764(30) \\
    & 16 & 30141 & 0.2857(54) \\
    & 20 & 8000 & 0.284(17) \\
\hline
2.5 & 4 & 15000 & 0.3990(16) \\
    & 6 & 51200 & 0.3815(14) \\
    & 8 & 21500 & 0.3653(32) \\
    & 12 & 37400 & 0.3621(39) \\
    & 16 & 33000 & 0.3618(58) \\
    & 20 & 4300 & 0.369(14) \\
\hline
3.0 & 4 & 12000 & 0.5427(17) \\
    & 6 & 51200 & 0.5193(14) \\
    & 8 & 15500 & 0.5113(44) \\
    & 12 & 48500 & 0.4958(36) \\
    & 16 & 19600 & 0.4941(74) \\
    & 20 & 10400 & 0.486(23) \\
\hline
3.5 & 4 & 15000 & 0.6769(16) \\
    & 6 & 51200 & 0.6495(14) \\
    & 8 & 17800 & 0.6367(37) \\
    & 12 & 57000 & 0.6244(34) \\
    & 16 & 18700 & 0.6105(66) \\
    & 20 & 4590 & 0.613(23) \\
\hline
4.5 & 4 & 15000 & 0.9370(14) \\
    & 6 & 51200 & 0.9067(14) \\
    & 8 & 14000 & 0.8884(41) \\
    & 12 & 63500 & 0.8849(33) \\
    & 16 & 33741 & 0.8671(52) \\
    & 20 & 8500 & 0.878(19) \\
\hline
8.0 & 4 & 12000 & 1.8221(23) \\
    & 6 & 51200 & 1.7904(13) \\
    & 8 & 14000 & 1.7619(40) \\
    & 12 & 44100 & 1.7492(41) \\
    & 16 & 12805 & 1.7258(90) \\
    & 20 & 10000 & 1.745(20) \\
\hline
\end{tabular}}
\label{tab:result}
\caption[a]{The measured $1/g^2(L)$ and the number of 
hybrid Monte Carlo trajectories used,
$N_{\rm traj.}$, at each lattice size $N$ and $\beta_L$.}
\end{table}
\endgroup

\begin{figure}[tb]
	\includegraphics[width=0.85\columnwidth]{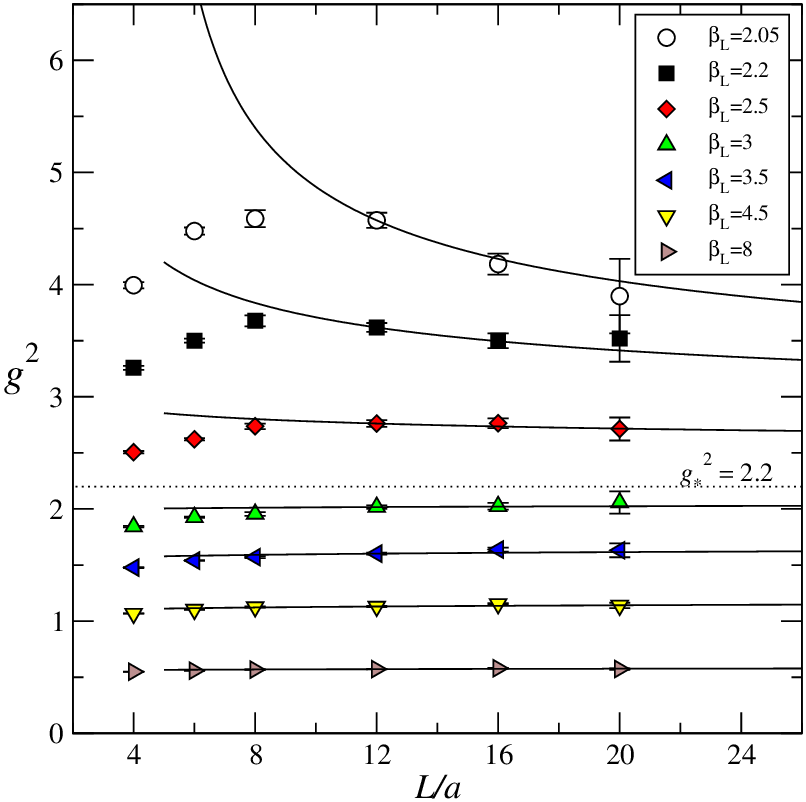}
	\caption{Lattice measurements of $g^2(L/a,\beta_L)$. 
        Continuous lines show $g^2(L)$ integrated from \eq{bfit1},
        constrained to go through lattice points at $L/a=12$.}
	\label{g2_su2a}
\end{figure}

\begin{figure}[tb]
	\includegraphics[width=0.85\columnwidth]{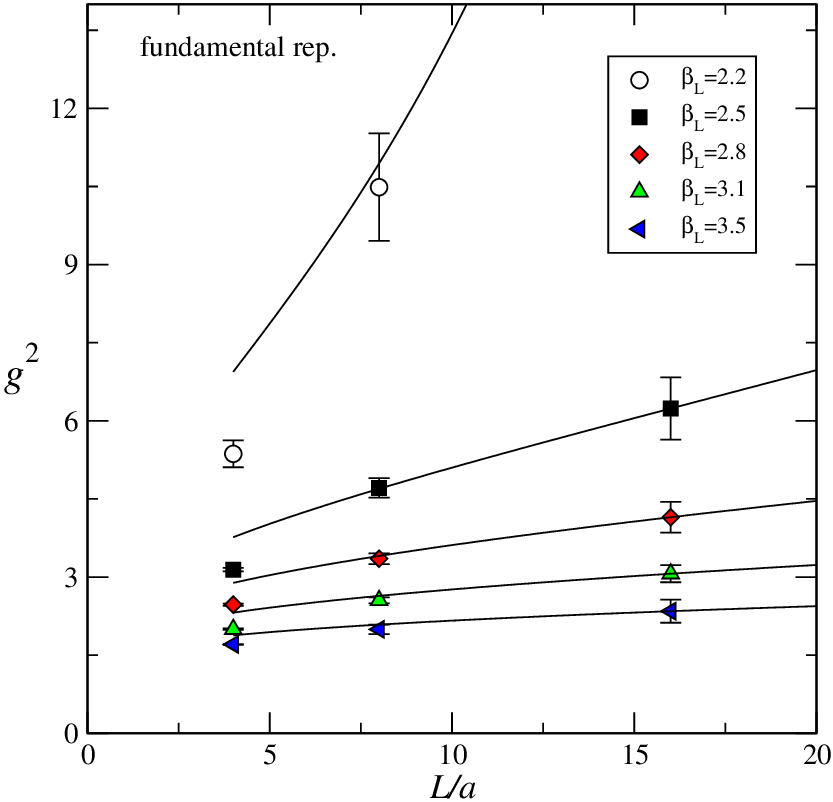}
	\caption{Small-statistic measurements of $g^2(L/a,\beta_L)$ 
          for 2 flavours of {\em fundamental} representation 
          fermions, to be contrasted with the adjoint representation
          case in Fig.~\ref{g2_su2a}.  The value
          of $g^2$ at $\beta_L=2.2$, $L/a=16$ is $26\pm 7$.
          Continuous lines show $g^2(L)$ integrated using 
          2-loop $\beta$-function, constrained to go through 
          lattice points at $L/a=16$.}
      \label{g2_su2}
\end{figure}

The measurements of $g^2(L/a,\beta_L)$ 
are done on lattices with volumes 
$V/a^4 = 4^4$ -- $20^4$, with the results shown in 
Fig.~\ref{g2_su2a} and tabulated in Table \ref{tab:result}.
We use a hybrid Monte Carlo (HMC) algorithm, with trajectory length
$\approx 1$; the table also shows the number of 
HMC trajectories for each point.  The measurement of
the coupling is done at the end of each 
trajectory.  The acceptance rate in the HMC algorithm
remains better than 84\% for all cases.
For comparison, we made also small-statistics study
of the coupling using two flavours of fundamental fermions.  These
results are shown in Fig.~\ref{g2_su2}.

What can we conclude from these figures?  
We remind that we can directly
compare physical lattice sizes only at fixed $\beta_L$;
different $\beta_L$ correspond to very different lattice spacings.
In the continuum $L$ is the sole physical scale, and $g^2(L)$
must be 
monotonic (or equivalently, $L$ is a function of $g^2$ through
dimensional transmutation).  Thus, the 
non-monotonic behaviour of data
seen at $\beta_L=2.05$ and $2.2$ is caused by finite
lattice spacing effects, i.e. too small $L/a$.
Hence, we do not include lattices with $L/a \le 8$ in
subsequent analysis.

Our most significant result is readily evident from
Fig.~\ref{g2_su2a}:  $g^2(L/a,\beta_L)$ is 
an increasing function of $L$ at $\beta_L \gsim 3$, ($g^2 \lsim 2.2$), 
but decreasing at $\beta_L \lsim 2.5$ ($g^2\gsim 2.2$) at large 
volumes. This is precisely the expected behaviour for a theory with
an infrared fixed point and markedly different from the asymptotically free QCD-like behaviour in Fig.~\ref{g2_su2}. In this case $g^2(L)$ is always increasing, and 
the larger $g^2$ is, the steeper the increase.
For the adjoint representation the observed 
fixed point $g^2_\ast \sim 2\ldots 3$ is significantly
below the two-loop result $7.9$.

Further analysis is needed in order to estimate the full $\beta$-function.
The standard method proceeds through the measurement of
step scaling functions: using lattices of sizes
$L$ and $s L$, where $s$ is a scale factor, we
measure $g^2(L/a,\beta_L)$ and $g^2(sL/a,\beta_L)$, thus 
obtaining a discrete step of the evolution of $g^2(L)$.
Now we can search for $\beta'_L$ so that 
$g^2(L/a,\beta'_L) = g^2(sL/a,\beta_L)$.  Repeating
these scaling and matching steps the evolution of $g^2$ over 
a very large range of scales can be covered.  
The continuum limit extrapolation is enabled by
redoing the analysis on different original $L/a$ 
\cite{Luscher:1992zx}.
This method has been successfully used, e.g. to determine
the running coupling in QCD \cite{DellaMorte:2004bc}.  

\begin{figure}[tb]
  \includegraphics[width=0.85\columnwidth]{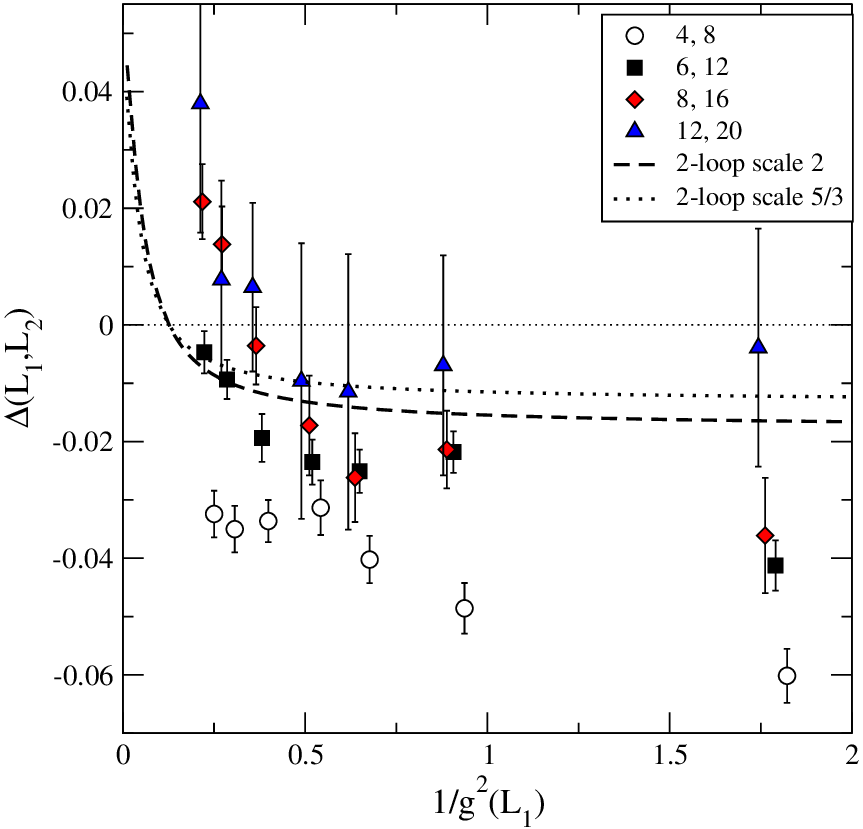}
  \caption{Step scaling $\Delta(L_1,L_2,\beta_L)$ for different pairs
    of volumes, see \eq{eq:discrete_beta}, plotted
    against $1/g^2(L_1,\beta_L)$.
    The scaling factor 
    is 2, except for the pair with largest volumes.
    Also shown is the 2-loop perturbative result for
    corresponding ratios of volumes.}
  \label{fig:step}
\end{figure}

In Fig.~\ref{fig:step} we show the step scaling using
the quantity
\begin{equation}
  \Delta(L_1,L_2,\beta_L) = \frac{1}{g^2(L_2,\beta_L)} 
  -\frac{1}{g^2(L_1,\beta_L)}.
  \label{eq:discrete_beta}
\end{equation}
We use scaling factor $s=L_2/L_1 = 2$, except for the 
largest volume $L/a=20$, where we do not have $L/a=10$
data avalable and use $s = 5/3$.  We also show the
2-loop scaling result computed from \eq{pert_beta}.

If finite lattice spacing effects were absent, all points with the
same $s$ should fall on a universal curve.  This is clearly not the
case; the small volume data lies significantly below the large volume
points.  We do not attempt to extrapolate the data to infinite volume
(continuum limit), because we do not trust the smallest volume data to
be close enough of continuum behaviour.  Nevertheless,
$\Delta(L_1,L_2,\beta_L)$ becomes negative at small $1/g^2$,
indicating the existence of the infrared fixed point.

The main reason for the difficulty of using the step scaling is that
the evolution of $g^2(L)$ is very slow, much slower than with
fundamental fermions.  The slow evolution is easily masked by finite
lattice spacing artifacts at small $L/a$, whereas the faster evolution
in fundamental fermion theory is rather easily observable.  Improved
actions can be expected to help significantly here.  The slow
evolution also increases the number of required recursive steps in the
step scaling method and hence statistical errors.

Due to the numerical problems with the step scaling function mentioned
above, we will therefore also consider the following alternative
method: we use the measured values of $g^2(L/a,\beta_L)$ on volumes
$12^4$--$20^4$, and fit a {\em $\beta$-function ansatz} to the data.
The ansatz includes the small-$g^2$ perturbative part and it should
have only a few tunable parameters.  The aim is to find an envelope of
continuum $\beta$-functions which are consistent with the
measurements.  The underlying assumption is that the lattice artifacts
are already sufficiently small at $L/a=12$.  This should be checked
with new simulations with improved actions in the future.

The simplest ansatz is the perturbatively motivated one $\beta(g) =
-b_0 g^3 - b_1 g^5 - b_2 g^7$.  Here $b_0$ and $b_1$ are fixed by the
perturbative $\beta$-function \eq{pert_beta}, and $b_2$ is a fit
parameter.  This function is actually consistent with the data,
but it is very likely too constrained giving unrealistically small
error range.  In order to have independent parameters for the value of
the fixed point coupling and the slope of the $\beta$-function at the
fixed point, we are led to the following ansatz:
\begin{equation}
   \beta(g) = -b_0 g^3 - b_1 g^5 + 
   (b_0 g_\ast^{3-\delta} +   b_1 g_\ast^{5-\delta})\,g^\delta .
  \label{bfit1} 
\end{equation}
Again $b_0$ and $b_1$ are fixed by the perturbation theory and
$g_\ast$ and $\delta$ are fit parameters which determine the fixed
point and the slope of the $\beta$-function around the fixed point.
The ansatz is naturally not of asymptotic form at small $g^2$, but
this does not matter for our purposes: the $g^\delta$-term turns out
to be negligible at small $g^2$, and the purpose of the ansatz is to
approximate the true $\beta$-function in the vicinity of the fixed
point.  In practice equivalent results are obtained using
perturbatively motivated ansatz $\ldots b_2 g^7 + b_3 g^9$; however,
\eq{bfit1} offers more freedom for the slope at $g^2 = g_\ast^2$.

We perform the fit as follows: integrating $\beta(g) = -L \dd g(L)/\dd L$
and constraining $g^2(L)$ to the lattice measurements at $L/a=12$ for
each $\beta_L$ gives us the continuous lines shown in
Fig.~\ref{g2_su2a}\footnote{Strictly speaking the fit is not optimal
  due to the constraint at $L/a=12$. Relaxing this constraint gives
  better $\chi^2$, at the cost of a more complicated fit procedure.
  However, because $L/a=12$ has smaller errors than larger volumes the
  effect is small.  }.  
The integrated $g^2(L/a,\beta_L)$ are compared with the lattice
measurements at $L/a=16$ and $20$, giving us a $\chi^2$ criterion for
the fit.  Good fits ($\chi^2/$d.o.f$\sim 1$) are obtained when
parameters vary from $(g_\ast^2,\delta) \approx (2.0,6)$ to $(3.2,15)$
in a narrow region.  Thus, the parameters are strongly correlated.  
The slope of the $\beta$-function at the fixed point is a universal
quantity, but  it is only weakly
constrained: 
\begin{equation}
  \left[g \frac{\dd \beta(g)}{\dd g}\right]_{g^2=g_\ast^2} 
  = 0.12 \ldots 0.42\,.
\end{equation}

We take $g_\ast^2 = 2.2$, $\delta = 7$ to be our benchmark value, which
is motivated by the fact that $\delta = 7$ is the order of the NNLO
perturbative $\beta$-function.  These lines are shown in Fig.~\ref{g2_su2a}.
However, we by no means imply that the
fitted value gives the perturbative result; it only quantifies all
unknown contributions.  If we fix $\delta = 7$, $g_\ast^2$ can vary
between 2.05 and 2.4.

\begin{figure}[tb]
 	\includegraphics[width=0.85\columnwidth]{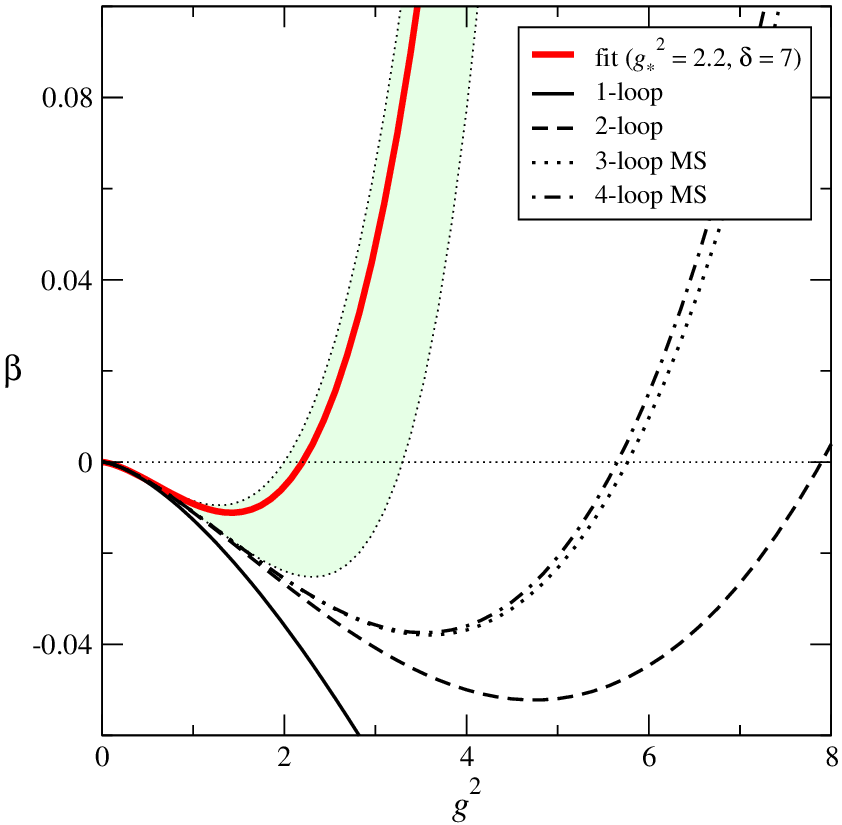}
 	\caption{The $\beta$-function obtained from
           \eq{bfit1}, with $g^2_\ast=2.2$. 
           The shaded area shows the estimated error range
           of the fit. Shown are also universal 
           perturbative one- and two-loop
           $\beta$-functions, together with the
           three- and four-loop results in the MS-scheme 
           \cite{vanRitbergen:1997va}.  Because of the
           different scheme, these are not directly
           comparable with the lattice (SF-scheme) results.}
 	\label{fig:beta}
 \end{figure}

The result of the fitting procedure is shown in Fig.~\ref{fig:beta},
together with the error band formed by the variation of the parameters
in the region where acceptable fit is obtained.
While the error band is wide, the qualitative features
agree with the corresponding step scaling functions 
calculated from the measurements; both methods indicate
positive $\beta(g)$ at large $g^2$.
We also note that the two-loop perturbative $\beta$-function alone fits
the measurements well at $\beta_L \gsim 3$ ($g^2 \lsim 2$).

Fig.~\ref{fig:beta} also shows the three- and four-loop MS-scheme results
\cite{vanRitbergen:1997va}.  Because of the different scheme, these
are not directly comparable with the SF-scheme results obtained in our
simulations --- for example, the value of $g_\ast^2$ is scheme
dependent.  However, it is interesting that also in the MS-scheme the
fixed point is at substantially smaller $g_\ast^2$ than indicated by
the universal two-loop calculation.\footnote{We note that the small
  difference between three- and four-loop results in MS-scheme is a numerical
  accident: varying the number of colours or the number of flavours
  typically gives a four-loop correction of roughly similar magnitude than the
  three-loop correction. Thus, we do not expect higher orders to be small in our case
  either.}




\begin{figure}[tb]
	\includegraphics[width=0.85\columnwidth]{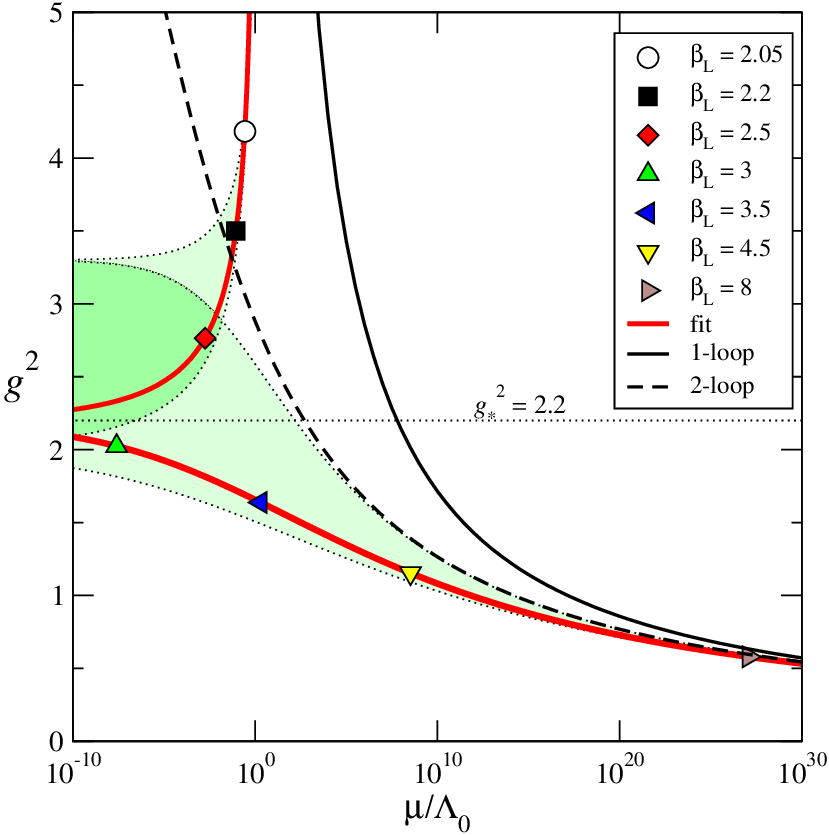}
	\caption{$g^2(\mu=1/L)$ determined from \eq{bfit1},
          together with perturbative $g^2$.  
          Scale $\Lambda_0$ is determined so that $\mu=\Lambda_0$
          is at the IR Landau pole
          of the 
          asymptotically free
          one-loop coupling 
          and at the UV Landau pole of the measured non-asymptotically free
          branch ($g^2 > g_\ast^2$).
          For illustration, the $L/a=16$ lattice points have been placed on
          the benchmark curve at the corresponding $g^2$-values, indicating
          the relative scale hierarchy between simulations.}
	\label{fig:g2}
\end{figure}

Finally, the fitted $\beta$-function can be integrated to obtain the coupling
$g^2$. This is shown in Fig.~\ref{fig:g2} as a function of $\mu =
1/L$.  The result is split into two branches: the asymptotically free
one with $g^2 < g_\ast^2$, and the not asymptotically free with
$g^2 > g_\ast^2$.  These branches 
are disconnected: when the $\beta$-function is integrated to obtain
$g^2(\mu)$, the integration constants are independent on both sides
of the fixed point. Thus, the scaling between the $\mu$-values between
the branches is arbitrary in Fig.~\ref{fig:g2}.
%

Finally, we note that the two-loop perturbative $\beta$-function
alone gives us an acceptable fit if we ignore the data at $\beta_L\le 2.5$
($g^2 \gsim 2.7$).  Indeed, in that case it is possible to fit a 
$\beta$-function which does not feature an IR fixed point.  Thus, the
evidence for the fixed point comes from the data at $\beta_L=2.05$
and $2.2$.  These are at relatively strong bare lattice coupling,
and one can expect large finite $a$ effects.  This can be clearly seen at
small $L/a$ in Fig.~\ref{g2_su2a}.  
Nevertheless, as $L/a\rightarrow\infty$ at fixed $\beta_L$, the scale
hierarchy between the physically interesting scale $L$ and the lattice
spacing $a$ increases, and the evolution of the coupling constant
approaches the continuum one.

\section{Conclusions and Outlook}

Let us now discuss the interpretation of these results. We note that
the observed range for the critical coupling, $g_\ast^2\sim 2.0 $ --
$3.2$ is significantly below the perturbative estimate $g_\ast^2\sim
8$ from \eq{pert_beta}.  Similar behaviour has been observed in SU(3)
gauge theory with 2-index symmetric fermions \cite{Shamir:2008pb}. The
obvious question is why the perturbative two-loop $\beta$-function
fails although the nonperturbative analysis implies that $\alpha_\ast
= g_\ast^2/4\pi$ is small. Since the perturbative determination of the
fixed point is based on the competing effects from one- and two-loop
orders, it is reasonable to expect that higher loop orders can
contribute with similar magnitude and one needs nonperturbative
analysis.  This is seen in the MS-scheme results in Fig.~\ref{fig:beta}.
Beyond two loops the value of $g_\ast^2$ is scheme dependent;
our result is in SF scheme. Comparison of the nonperturbative result
and the perturbative curves in Fig.\,\ref{fig:beta} demonstrates that
even if the actual value of the coupling at the fixed point is small,
the nonperturbative effects can be large. 

When we combine the results of the analysis here with the spectrum
measurements of ref.~\cite{Hietanen:2008mr}, a consistent picture
emerges. 
%
In the whole parameter range where 
simulations with massless fermions are possible the infrared
behaviour is controlled by the conformally invariant 
fixed point $g_\ast^2$ and there is no 
chiral symmetry breaking.  If $g^2 < g_\ast^2$, the
theory is asymptotically free, otherwise not.  The
conformal invariance can be explicitly broken by e.g. adding
a small fermion mass, which can lead to a ``walking'' type
evolution. 

As discussed above, for the lattice action used here
it has been observed that it is not possible to take the limit 
$m_Q\rightarrow 0$ when $\beta_L \lsim 2$.
Thus, the $\beta_L = 2.05$ -points in Fig.~\ref{fig:g2} are close
to the smallest $\beta_L$ where the conformal behaviour 
can be studied.
This limitation is a non-universal lattice artifact, and may differ for 
some other fermion  discretisation.
This
motivates a study with fully $O(a)$ improved lattice formalism.

{\em Acknowledgement:~} KR acknowledges the support of Academy of
Finland grant 114371. AH acknowledges the support by the NFS under
grant number PHY-055375 and DOE grant under contract
DE-FG02-01ER41172. The simulations were performed at the Center for
Scientific Computing (CSC), Finland, and the total computing effort
amounts to $\sim 10^{18}$~flop.


\end{document}